\documentclass[prl,twocolumn,floatfix,showpacs ]{revtex4-1}
\UseRawInputEncoding
\usepackage{amsmath}
\usepackage{amssymb}
\usepackage{graphicx}
\usepackage{dcolumn}
\usepackage{bm}
\usepackage[colorlinks=true,linkcolor=blue,anchorcolor=blue, citecolor=cyan,urlcolor=cyan]{hyperref}
\usepackage[mathlines]{lineno}
\usepackage{ulem}
\usepackage{pifont}
\usepackage{epstopdf}
\begin{document}
\title{Symmetry-breaking induced transition among net-zero-magnetization magnets}
\author{San-Dong Guo}
\email{sandongyuwang@163.com}
\affiliation{School of Electronic Engineering, Xi'an University of Posts and Telecommunications, Xi'an 710121, China}
\author{Xiao-Shu Guo}
\affiliation{School of Electronic Engineering, Xi'an University of Posts and Telecommunications, Xi'an 710121, China}
\author{Guangzhao Wang}
\affiliation{Key Laboratory of Extraordinary Bond Engineering and Advanced Materials Technology of Chongqing, School of Electronic Information Engineering, Yangtze Normal University, Chongqing 408100, China}

\begin{abstract}
Net-zero-magnetization magnets have garnered intensive research attention due to their  ultradense and ultrafast potential.
 In terms of the symmetric classification of connecting magnetic atoms with opposite spin polarization, the net-zero-magnetization magnets mainly include  $PT$-antiferromagnet (the joint symmetry ($PT$) of space inversion symmetry ($P$) and time-reversal symmetry ($T$)), altermagnet and fully-compensated ferrimagnet. Studying transitions among net-zero-magnetization magnets is essentially the  research on symmetry breaking, which can also clearly reveal the transformation of spin-splitting symmetry. Symmetry breaking can be achieved through methods such as Janus engineering, isovalent alloying, and external electric field. Here, we start from a parent $PT$-antiferromagnet that simultaneously possesses both $P$ and rotational/mirror symmetries to induce altermagnet and fully-compensated ferrimagnet. Based on first-principles calculations, the proposed transitions can be verified in $PT$-antiferromagnet $\mathrm{CrC_2S_6}$ monolayer.  By Janus engineering  and  isovalent alloying, $\mathrm{CrC_2S_6}$  can change into altermagnetic $\mathrm{CrC_2S_3Se_3}$   and fully-compensated ferrimagnetic $\mathrm{CrMoC_2S_6}$.  The $\mathrm{CrC_2S_3Se_3}$ can also become fully-compensated ferrimagnetic   $\mathrm{CrMoC_2S_3Se_3}$ by  isovalent alloying. Our work provides a clear and intuitive example to explain the transitions among net-zero-magnetization magnets, which can inspire more research on net-zero-magnetization magnets.

\end{abstract}
\maketitle
\textcolor[rgb]{0.00,0.00,1.00}{\textbf{Introduction.---}}
Ferromagnets are one of the most important materials in spintronics\cite{o1}. The spontaneous magnetization of ferromagnets makes them a stable magnetic storage medium in spintronics. The magnetization direction of ferromagnets can be flipped by an external magnetic field or current, and they are widely used in magnetic random access memory (MRAM) and other storage devices. The ferromagnets can also be used to develop highly sensitive magnetic sensors and readout devices. Although ferromagnets have important applications in spintronics, they also have some limitations: stray field problems and power consumption issues\cite{k1,k1-1,k2}. Instead, contemporary interest in spintronics has shifted to net-zero-magnetization magnets. Compared with ferromagnets, net-zero-magnetization magnets offer more advantages for spintronic devices, particularly in terms of high storage density, robustness against external magnetic fields, and ultrafast writing speed, which are all due to their net-zero magnetic moment\cite{k1,k1-1,k2}.  $PT$-antiferromagnet (the joint symmetry ($PT$) of space inversion symmetry ($P$) and time-reversal symmetry ($T$)), altermagnet and fully-compensated ferrimagnet are typical net-zero-magnetization magnets\cite{k4,k5}.

 \begin{figure}[t]
    \centering
    \includegraphics[width=0.45\textwidth]{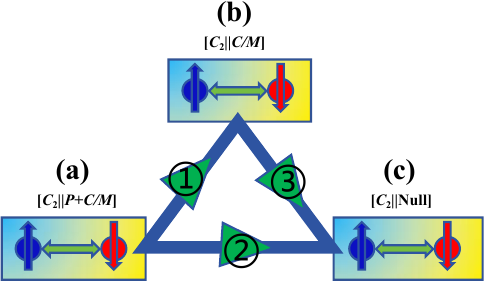}
    \caption{(Color online) The net-zero-magnetization magnet mainly includes $PT$-antiferromagnet (a), altermagnet (b) and fully-compensated ferrimagnet (c).  The magnetic atoms with opposite spins are connected by inversion symmetry ($P$),  rotation/mirror symmetry ($C/M$) and no symmetry, respectively. To study the transformation between them, the symmetric connection in $PT$-antiferromagnet is restricted to $P$ plus $C/M$.  By breaking symmetry with an external field, $PT$-antiferromagnet can be transformed into altermagnet  and fully-compensated ferrimagnet (\textcircled{1} and \textcircled{2}), and  altermagnet can also be transformed into fully-compensated ferrimagnet (\textcircled{3}).}\label{a}
\end{figure}

\begin{figure*}[t]
    \centering
    \includegraphics[width=0.75\textwidth]{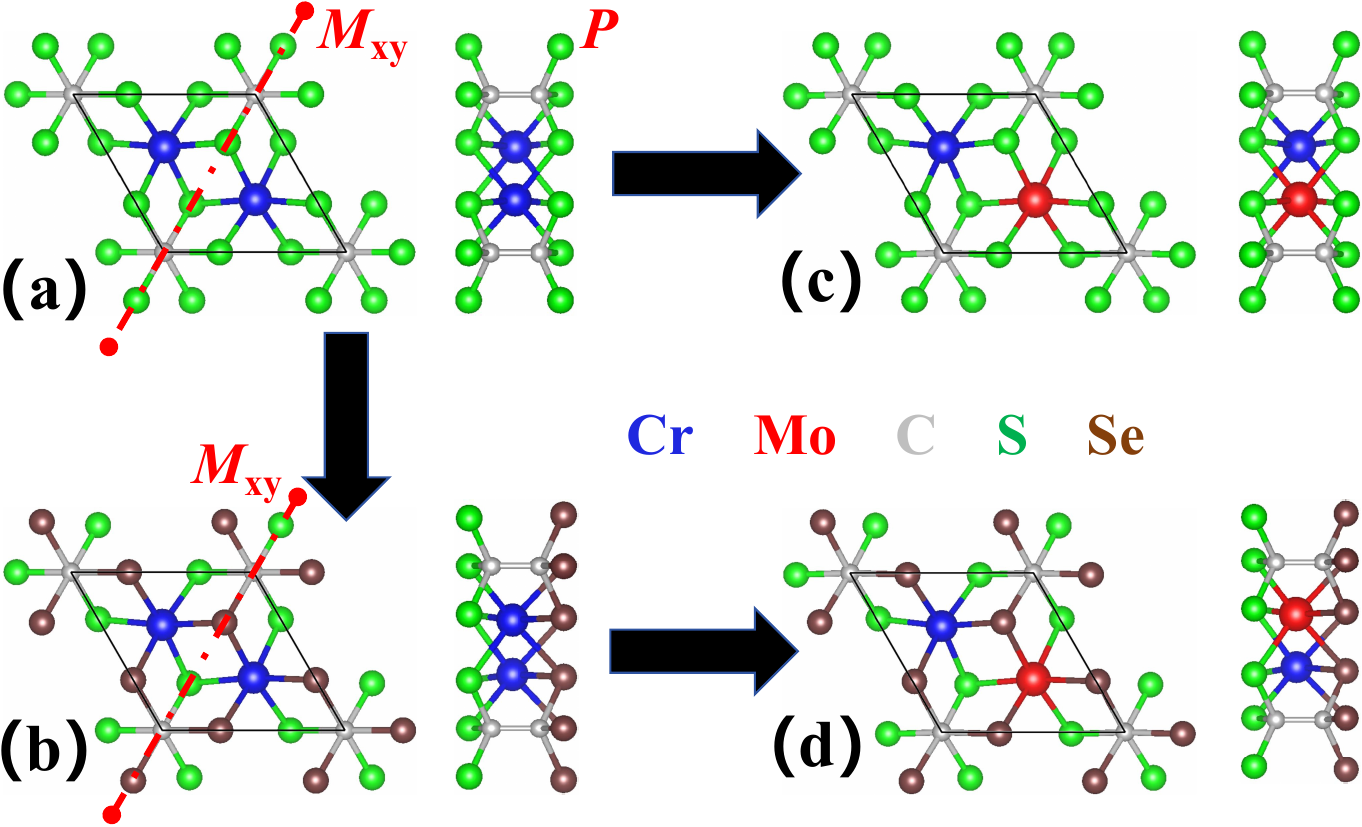}
    \caption{(Color online) Transitions among  crystal structures:  the $\mathrm{Cr_2C_2S_6}$ (a) can be a parent $PT$-antiferromagnet with $P$ and $M_{xy}$ symmetries. By Janus engineering, the $\mathrm{Cr_2C_2S_6}$ becomes  $\mathrm{Cr_2C_2S_3Se_3}$ (b)  with $M_{xy}$ symmetry as a  altermagnet.  By isovalent alloying,  the $\mathrm{Cr_2C_2S_6}$ and $\mathrm{Cr_2C_2S_3Se_3}$  can  become  fully-compensated ferrimagnetic  $\mathrm{CrMoC_2S_6}$ (c) and  $\mathrm{CrMoC_2S_3Se_3}$.   }\label{b}
\end{figure*}

Traditional antiferromagnets,  such as  $PT$-antiferromagnet, exhibit spin degeneracy, which limits many interesting physical phenomena and effects.
 The spin-splitting can be induced  by making the magnetic atoms with opposite spin polarization locating in the different environment\cite{gsd}.
That is to say, it is necessary to break the $P$ symmetry of the system\cite{qq1,qq2}, and $PT$-antiferromagnet can become altermagnet and fully-compensated ferrimagnet with spin-splitting.
Altermagnetism and fully-compensated ferrimagnetism not only share certain key properties with antiferromagnetism but also exhibit even more similarities with ferromagnetism.   Unlike traditional  antiferromagnet,  both altermagnet and fully-compensated ferrimagnet  can  give rise to anomalous Halll/Nernst effect  and magneto-optical Kerr effect\cite{k4,h13,f4}.

The altermagnet exhibits   momentum-dependent spin-splitting of $d$-, $g$-, or $i$-wave symmetry  in Brillouin zone (BZ) without the help of spin-orbital coupling (SOC)\cite{k4,k5,k511,k512,k513}. Experimentally and theoretically, several bulk and two-dimensional (2D) altermagnetic materials have been identified that exhibit momentum-dependent spin-splitting\cite{k4,h13,k60,k6,k7,k7-1,k7-2,k7-3}. Recently, twisted altermagnetism, utilizing one of the five 2D Bravais lattices, has been proposed in twisted magnetic van der Waals (vdW) bilayers\cite{k8,k80}. In these systems, an out-of-plane electric field can induce valley polarization due to valley-layer coupling\cite{k9,k10}. Additionally, an antiferroelectric altermagnet has been proposed, featuring the coexistence of antiferroelectricity and altermagnetism in a single material\cite{k7-3-2}.

Fully-compensated ferrimagnets constitute a distinctive category of ferrimagnetic materials, marked by their net-zero magnetization\cite{f1,f2,f3},  the spin-splitting  of which   occurs in the whole BZ with $s$-wave symmetry, being similar to the situation of ferromagnet.
Recently, the significance of 2D fully-compensated ferrimagnets has been increasingly recognized, expanding the realm of low-dimensional spintronic materials\cite{f4}. The net-zero magnetization of  fully-compensated ferrimagnetism  is due to  gap-guaranteed spin quantization in one spin channel\cite{f4}, not symmetry constraints.   In our previous works\cite{gsd,gsd0,gsd1,gsd2,gsd3,gsd4},   $\mathrm{CrMoC_2S_6}$, $\mathrm{V_2F_7Cl}$ and many Janus A-type antiferromagnetic (AFM)-ordering monolayers  are strictly 2D fully-compensated ferrimagnets, not antiferromagnets.

Intuitively and vividly elucidating the transitions among these net-zero-magnetization magnets will help to further reveal the connections of their physical effects and promote their applications in spintronic devices. Here, we start from a parent $PT$-antiferromagnet  to induce altermagnet and fully-compensated ferrimagnet. Based on first-principles calculations,  we take parent  $PT$-antiferromagnet $\mathrm{CrC_2S_6}$ monolayer as example to achieve  altermagnetic $\mathrm{CrC_2S_3Se_3}$   and fully-compensated ferrimagnetic $\mathrm{CrMoC_2S_6}$/$\mathrm{CrMoC_2S_3Se_3}$ by Janus engineering  and  isovalent alloying.

\begin{figure}[t]
    \centering
    \includegraphics[width=0.45\textwidth]{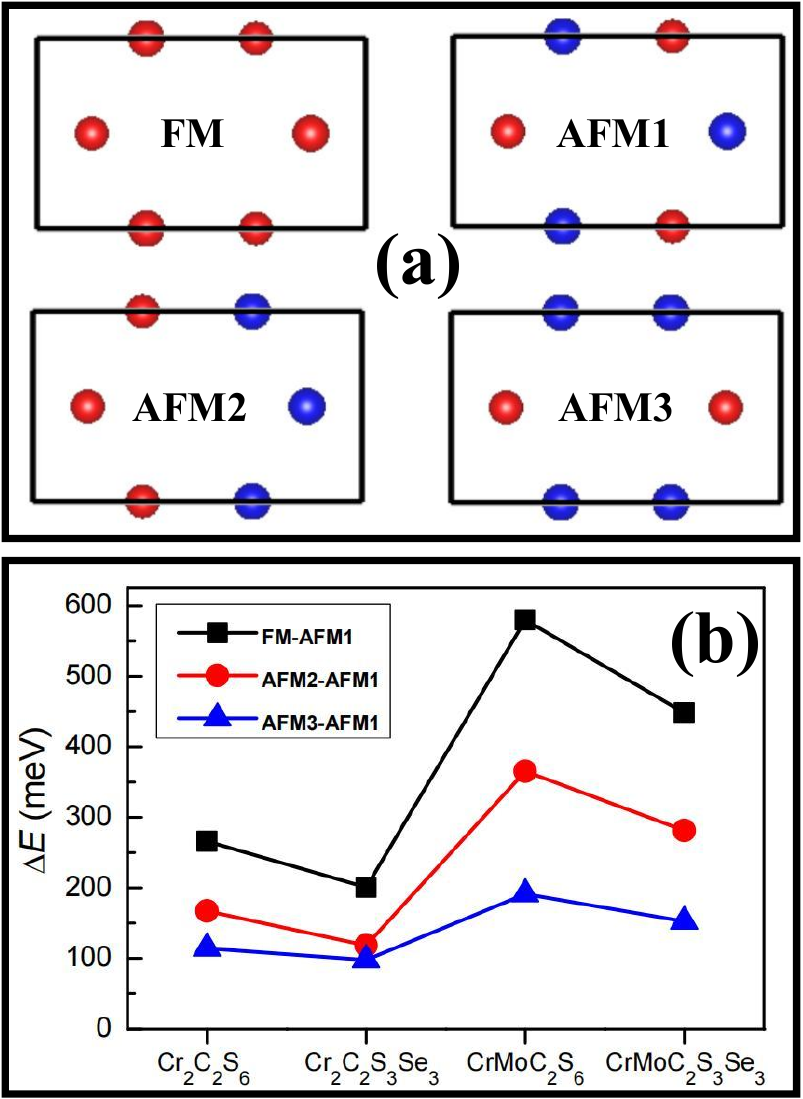}
     \caption{(Color online) (a): four magnetic configurations,  including FM, AFM1, AFM2 and AFM3, and the red and blue balls represent spin-up and spin-down atoms, respectively. (b):the energy differences between  FM/AFM2/AFM3 and AFM1 orderings for monolayer $\mathrm{Cr_2C_2S_6}$, $\mathrm{Cr_2C_2S_3Se_3}$, $\mathrm{CrMoC_2S_6}$ and  $\mathrm{CrMoC_2S_3Se_3}$. }\label{c}
\end{figure}
\begin{figure*}[t]
    \centering
    \includegraphics[width=0.75\textwidth]{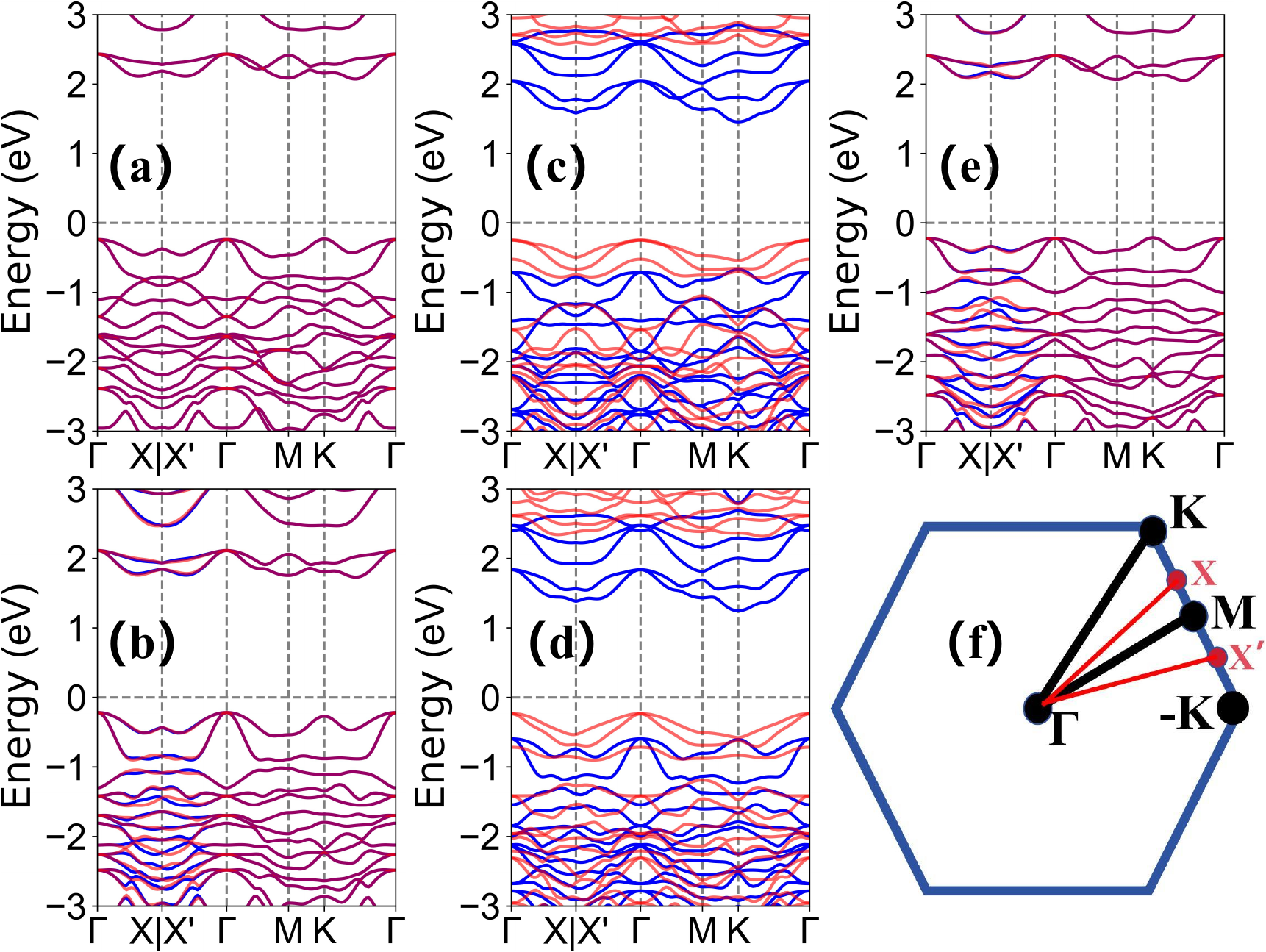}
     \caption{(Color online) The spin-polarized  energy band structures of  $\mathrm{Cr_2C_2S_6}$ (a), $\mathrm{Cr_2C_2S_3Se_3}$ (b), $\mathrm{CrMoC_2S_6}$ (c),  $\mathrm{CrMoC_2S_3Se_3}$ (d)  and $\mathrm{Cr_2C_2S_6}$ at  $E$=0.30$\mathrm{V/{\AA}}$ (e)  along with BZ   with high symmetry points (f).   In (a, b, c, d, e), the spin-up and spin-down channels are depicted in blue and red, and the  purple color means spin degeneracy. }\label{d}
\end{figure*}

\textcolor[rgb]{0.00,0.00,1.00}{\textbf{Transition among  net-zero-magnetization magnets.---}}
From a symmetric point of view\cite{k4,k5},  the net-zero-magnetization collinear magnets mainly include $PT$-antiferromagnet, altermagnet and fully-compensated ferrimagnet (see \autoref{a}), and   their magnetic sublattices can be connected  by the [$C_2$$\parallel$$P$],   [$C_2$$\parallel$$C/M$] and  the [$C_2$$\parallel$Null], where  $C_2$ and $C$/$M$ are the two-fold rotation perpendicular to the spin axis in spin space, and  rotation/mirror symmetry in lattice space. To study the transformation between them, the symmetric connection of  two sublattices  in $PT$-antiferromagnet is restricted to $P$ plus $C/M$.
 The $PT$-antiferromagnets are spin degenerate due to $PT$ symmetry ($E_{\uparrow}(k)$=$PT$$E_{\uparrow}(k)$= $E_{\downarrow}(k)$), while the altermagnet and fully-compensated ferrimagnet show momentum-dependent (such  as $d$-wave, $g$-wave  and $i$-wave symmetry) and global ($s$-wave symmetry) spin-splitting, respectively.

 If  $PT$-antiferromagnet is transformed into altermagnet through symmetry breaking, it should possess not only $P$ symmetry but also $C/M$ symmetry.
 By breaking $P$ symmetry while preserving $C/M$ symmetry, the transformation from $PT$-antiferromagnet to altermagnet can be achieved (\textcircled{1} of \autoref{a}). To transform from $PT$-antiferromagnet and altermagnet to fully-compensated ferrimagnet, it is only necessary to break both $P$ and $C/M$ symmetries (\textcircled{2} and \textcircled{3}  of \autoref{a}).

In fact, net-zero-magnetization magnets can be transformed into each other through stacking vdW bilayer engineering. Initially, we take magnetic  monolayer as the basic building unit  to build the bilayer, where the upper and lower layers can be connected by   [$C_2$$\parallel$$P$] symmetry.
By breaking horizontal mirror symmetry of $PT$-bilayer antiferromagnet through a sliding operation, an out-of-plane built-in electric field can be induced\cite{qq3}, which in turn can produce  a global spin-splitting, leading to a fully-compensated ferrimagnet.  Through twisted-angle engineering\cite{k8,k80}, the upper and lower layers of $PT$-bilayer antiferromagnet can be connected through rotational symmetry, which can lead to an altermagnet.  If an out-of-plane external electric field is applied, $PT$-bilayer antiferromagnet and twisted altermagnet  can both become fully-compensated ferrimagnet\cite{k9,gsd1,gsd2,gsd3}. These reverse processes can be achieved through opposite operations.

Here,  we focus on symmetry-breaking induced transition among  net-zero-magnetization magnets.  In other words, starting from $PT$-antiferromagnet with the highest symmetry, altermagnet and fully-compensated ferrimagnet  are induced through symmetry reduction. Firstly, we search for $PT$-antiferromagnet with $C/M$ symmetry by magnetic point group, which can be achieved by high-throughput screening,  and then identify more candidates.  Secondly,  altermagnet and fully-compensated ferrimagnet can be achieved  by breaking the corresponding symmetries through electric field, Janus engineering, alloying, defects, and so on.
The electric field is a volatile regulation, while the Janus engineering, alloying and  defects are  non-volatile construction.

\textcolor[rgb]{0.00,0.00,1.00}{\textbf{Computational detail.---}}
The  spin-polarized  first-principles calculations  are carried out within density functional theory (DFT) \cite{1},  as implemented in Vienna ab initio simulation package (VASP)\cite{pv1,pv2,pv3} by using the projector augmented-wave (PAW) method. We use  generalized gradient approximation (GGA) of  Perdew, Burke, and Ernzerhof (PBE)\cite{pbe} as the exchange-correlation functional. The kinetic energy cutoff  of 500 eV,  total energy  convergence criterion of  $10^{-8}$ eV, and  force convergence criterion of  0.0001 $\mathrm{eV.{\AA}^{-1}}$ are adopted  to obtain the reliable results.
The  Hubbard correction is added  with $U$=3.00 eV\cite{f7-1} for $d$-orbitals of both Cr and Mo atoms within the
rotationally invariant approach proposed by Dudarev et al\cite{du}.
 To avoid interlayer interactions, we use a slab model with a vacuum thickness of more than 15 $\mathrm{{\AA}}$ along $z$ direction.
The  BZ is sampled  with a 13$\times$13$\times$1 Monkhorst-Pack $k$-point meshes  for structure relaxation and electronic structure calculations.

The phonon dispersion spectrums  are obtained  within finite displacement method by using  3$\times$3$\times$1 supercell, as implemented in the  Phonopy code\cite{pv5}.  The ab initio
molecular dynamics (AIMD) simulations   are performed with a 3$\times$3$\times$1 supercell  for more than
8000 fs with a time step of 1 fs.  The elastic stiffness tensor  $C_{ij}$   are calculated by using strain-stress relationship, and they are  renormalized by   $C^{2D}_{ij}$=$L_z$$C^{3D}_{ij}$ with the $L_z$ being  the length of unit cell along $z$ direction.  The magnetic orientation is determined  by calculating  magnetic anisotropy energy (MAE): $E_{MAE}=E^{||}_{SOC}-E^{\perp}_{SOC}$, where $||$ and $\perp$  mean that spins lie in
the plane and out-of-plane. The Berry curvatures can be calculated from
wave functions  based on Fukui's
method\cite{bm} by using the VASPBERRY code\cite{bm1,bm2,bm3}.

\textcolor[rgb]{0.00,0.00,1.00}{\textbf{Material realization.---}}
Here, we take monolayer $\mathrm{Cr_2C_2S_6}$ as a prototype example to illustrate symmetry-breaking induced transition among  net-zero-magnetization magnets.
As shown in \autoref{b} (a), the Cr atoms in $\mathrm{Cr_2C_2S_6}$ are surrounded by six S
atoms, forming a honeycomb lattice, and two $\mathrm{CrS_3}$ moieties are connected by two C atoms.
 The $\mathrm{Cr_2C_2S_6}$ crystallizes in the  $P\bar{3}1m$ space group (No.162),  possessing $P$ and $M_{xy}$ symmetries.
 By replacing the top S layer with Se element in $\mathrm{Cr_2C_2S_6}$, the Janus monolayer $\mathrm{Cr_2C_2S_3Se_3}$ is obtained (\autoref{b} (b)), which has the symmetry of $P31m$ (No.157). The $\mathrm{Cr_2C_2S_3Se_3}$  lacks $P$ symmetry but still maintains $M_{xy}$ symmetry, which provides the fundamental condition for achieving altermagnet.  The $\mathrm{CrMoC_2S_6}$/$\mathrm{CrMoC_2S_3Se_3}$ (see \autoref{b} (c)/(d)) can be obtained by substituting one Cr of  $\mathrm{Cr_2C_2S_6}$/$\mathrm{Cr_2C_2S_3Se_3}$ with Mo via isovalent alloying, which crystallizes in the  $P312$/$P3$ space group (No.149/No.143),  lacking both $P$ and $M_{xy}$ symmetries as fully-compensated ferrimagnet.

To determine the magnetic  ground state of the four monolayers,
four magnetic configurations, including ferromagnetic (FM), AFM1, AFM2 and AFM3 orderings,  have
been constructed, as displayed in \autoref{c} (a).  The energy differences between  FM/AFM2/AFM3 and AFM1 orderings  are plotted in  \autoref{c} (b).
For all cases, the energy difference is positive, indicating that AFM1 ordering is the magnetic ground state of the four monolayers. The AFM1 ordering, along with the same magnetic ordering of  the four monolayers, provides the fundamental condition and convenience for studying the transformation among  net-zero-magnetization magnets.
Within AFM1 ordering, the  optimized  equilibrium lattice constants are  $a$=$b$=5.636, 5.808, 5.714 and 5.883 $\mathrm{{\AA}}$  for monolayer $\mathrm{Cr_2C_2S_6}$, $\mathrm{Cr_2C_2S_3Se_3}$, $\mathrm{CrMoC_2S_6}$ and  $\mathrm{CrMoC_2S_3Se_3}$.

To explore the stabilities of four monolayers,
the phonon dispersion,  AIMD  and elastic constants are carried out by using GGA+$U$ for AFM1ordering. The calculated phonon spectrums of four monolayers  are plotted in FIG.S1\cite{bc},  and they show no obvious imaginary frequencies,  indicating their dynamic stabilities.
The  total
energy as a function of simulation time and  the final crystal structure at 8 ps are shown in FIG.S2\cite{bc} using AIMD at 300 K,  which show  that the thermal-induced  energy fluctuations and  changes in geometry are small, indicating that the four monolayers have  good thermal stabilities at 300 K. For monolayer $\mathrm{Cr_2C_2S_6}$/$\mathrm{Cr_2C_2S_3Se_3}$/$\mathrm{CrMoC_2S_6}$/$\mathrm{CrMoC_2S_3Se_3}$, the independent elastic constants are $C_{11}$=98.83/90.16/91.69/84.42  $\mathrm{Nm^{-1}}$, $C_{12}$=29.56/28.31/34.57/32.36 $\mathrm{Nm^{-1}}$ ,  and they all  satisfy the  Born  criteria of mechanical stability\cite{ela}: $C_{11}>0$ and $C_{11}-C_{12}>0$,  confirming thier mechanical stabilities.

Without including SOC, the spin-polarized  energy band structures of  $\mathrm{Cr_2C_2S_6}$, $\mathrm{Cr_2C_2S_3Se_3}$, $\mathrm{CrMoC_2S_6}$ and   $\mathrm{CrMoC_2S_3Se_3}$  are plotted in \autoref{d} (a, b, c, d). To clearly observe the altermagnetic spin-splitting, we have added the $\Gamma$-X$\mid$X'-$\Gamma$ path, which is shown in  \autoref{d} (f). Due to $PT$ symmetry, $\mathrm{Cr_2C_2S_6}$ exhibits spin degeneracy throughout the entire BZ, and it is an indirect bandgap semiconductor with a gap value of 2.31 eV.  The total magnetic moment of $\mathrm{Cr_2C_2S_6}$ is strictly equal to 0  $\mu_B$, with the magnetic moments of the two magnetic atoms being 2.983 $\mu_B$ and -2.983 $\mu_B$, respectively.

By  Janus engineering,   a built-in electric field (about 1.02 $\mathrm{V/{\AA}}$  in FIG.S3\cite{bc}) in $\mathrm{Cr_2C_2S_3Se_3}$ breaks the $PT$ symmetry, but preserves the mirror symmetry,  giving rise to altermagnetic spin-splitting in energy bands.
 In the absence
of the SOC, the symmetry of $\mathrm{Cr_2C_2S_3Se_3}$ possesses a sixfold rotation,   leading to  spin-splitting of $i$-wave symmetry. The $\mathrm{Cr_2C_2S_3Se_3}$  is also  an indirect bandgap semiconductor  of  1.947 eV. The total magnetic moment of $\mathrm{Cr_2C_2S_3Se_3}$ is strictly equal to 0  $\mu_B$,  and the magnetic moments of  two Cr atoms are  3.067 $\mu_B$ and -3.067 $\mu_B$, respectively.
 A built-in electric field can be equivalently replaced by external electric field $E$, and the spin-polarized  energy band structures of $\mathrm{Cr_2C_2S_6}$ at  $E$=0.30$\mathrm{V/{\AA}}$  are plotted in \autoref{d} (e), which shows a similar  altermagnetic spin-splitting with that of  $\mathrm{Cr_2C_2S_3Se_3}$. Its total magnetic moment  is  still equal to 0  $\mu_B$,  and the magnetic moments of  two Cr atoms are  2.988 $\mu_B$ and -2.988 $\mu_B$, respectively.
When the direction of the electric field is reversed, the order of spin-splitting also reverses (see FIG.S4\cite{bc}), which is feasible for tuning the spin current in spintronic devices.

\begin{figure}[t]
    \centering
    \includegraphics[width=0.45\textwidth]{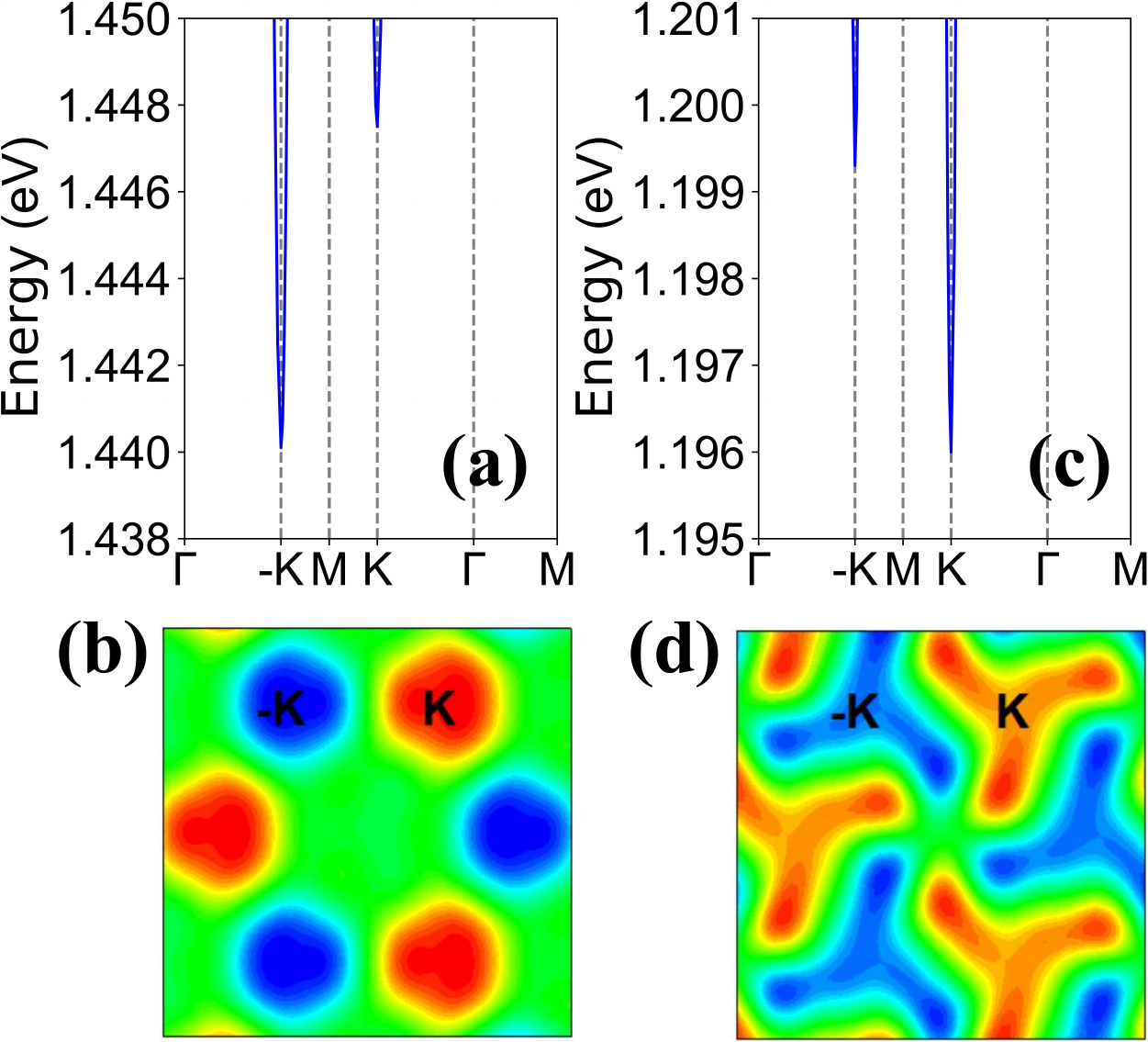}
     \caption{(Color online)  The enlarged energy band structures of conduction bands near the Fermi energy level and the distribution of Berry curvatures of  $\mathrm{CrMoC_2S_6}$ (a, b) and  $\mathrm{CrMoC_2S_3Se_3}$ (c, d)  within SOC. In (b, d), the green and red colors represent positive and negative values, respectively. }\label{e}
\end{figure}

 By isovalent alloying in  $\mathrm{Cr_2C_2S_6}$ and  $\mathrm{Cr_2C_2S_3Se_3}$, both $P$ and $M_{xy}$ symmetries can be broken, which leads to the magnetic atoms with opposite spins being connected asymmetrically in $\mathrm{CrMoC_2S_6}$  and  $\mathrm{CrMoC_2S_3Se_3}$,  producing fully-compensated ferrimagnetic spin-splitting of $s$-wave symmetry.   The large spin-splitting around the
Fermi energy level  is due to $d$ orbital mismatch between Cr and Mo magnetic atoms\cite{gsd}.
 Both $\mathrm{CrMoC_2S_6}$  and  $\mathrm{CrMoC_2S_3Se_3}$ are indirect bandgap semiconductors, with bandgaps of 1.702 eV  and 1.479 eV, respectively.
The total magnetic moment of  $\mathrm{CrMoC_2S_6}$/$\mathrm{CrMoC_2S_3Se_3}$ is strictly equal to 0  $\mu_B$, and  the magnetic moments of the Cr and Mo atoms are   2.892/2.985 $\mu_B$ and   -2.355/-2.406 $\mu_B$. The different absolute values of the magnetic moments of Cr and Mo atoms with opposite spin polarization are due to their asymmetric connections. From  $\mathrm{CrMoC_2S_6}$ to $\mathrm{CrMoC_2S_3Se_3}$,  the  $\Gamma$-X and X'-$\Gamma$ paths will become inequivalent due to the symmetry reduction,  leading to the lack of symmetry in the band structure (see FIG.S5\cite{bc}).

Achieving valley polarization and the anomalous valley Hall effect (AVHE) in net-zero-magnetization magnets is of great significance for the development of valleytronics devices.  Valley polarization can exist in hexagonal $PT$-antiferromagnet\cite{gsd1,gsd2}. To achieve AVHE, there should be spin-splitting, which requires breaking the $PT$ symmetry, thus transforming into fully-compensated ferrimagnet. For twisted altermagnet, valley polarization can be induced by an external out-of-plane electric field\cite{k9,k10}, which in fact transforms it into fully-compensated ferrimagnet. Therefore, fully-compensated ferrimagnet is a natural material for realizing the AVHE.
Firstly,  we determine the magnetization directions of  fully-compensated ferrimagnetic $\mathrm{CrMoC_2S_6}$ and $\mathrm{CrMoC_2S_3Se_3}$ by calculating MAE, because only  out-of-plane magnetization  can produce spontaneous valley polarization\cite{gsd1,gsd2}. The calculated MAE is 148/209$\mathrm{\mu eV}$/unit cell for $\mathrm{CrMoC_2S_6}$/$\mathrm{CrMoC_2S_3Se_3}$, which  means the out-of-plane easy magnetization axis.

The energy band structures of  $\mathrm{CrMoC_2S_6}$/$\mathrm{CrMoC_2S_3Se_3}$  with SOC are plotted in FIG.S6\cite{bc},  and the enlarged energy band structures of conduction bands near the Fermi energy level and the distribution of Berry curvatures ($\Omega(k)$) are shown in \autoref{e}.
Based on  \autoref{e} (a, c), it is clearly seen  that there exists valley polarization.
 The energy of K valley of $\mathrm{CrMoC_2S_6}$
is higher than one of -K valley  with the valley splitting of  7.4  meV ($|\Delta E_C|$=$|E_{K}^C-E_{-K}^C|$). However,  for $\mathrm{CrMoC_2S_3Se_3}$,
the energy of -K valley
is higher than one of  K valley  with the $|\Delta E_C|$ of  3.3  meV.
It is clearly seen that the Berry curvatures of the -K and K valleys have opposite signs.
 When  a longitudinal in-plane electric field $E_{\parallel}$ is applied,
the Bloch carriers will acquire an anomalous transverse
velocity $v_{\bot}$$\sim$$E_{\parallel}\times\Omega(k)$\cite{qq4}. When
 the Fermi energy level is shifted between the -K and K valleys in the conduction band of $\mathrm{CrMoC_2S_6}$/$\mathrm{CrMoC_2S_3Se_3}$, the spin-up carriers from -K/K valley will
accumulate along one edge of the sample,  giving rise to the AVHE (FIG.S7\cite{bc}).

\textcolor[rgb]{0.00,0.00,1.00}{\textbf{Conclusion.---}}
 In summary, we present the transition  among net-zero-magnetization magnets,  including  $PT$-antiferromagnet, altermagnet and fully-compensated ferrimagnet.
From the parent material $PT$-antiferromagnet, altermagnet and fully-compensated ferrimagnet can be derived through symmetry breaking. Specifically, the parent $PT$-antiferromagnet possesses initial symmetries $P$ and $C/M$. Through the process of symmetry breaking, materials with different symmetry of spin-splitting,  altermagnet and fully-compensated ferrimagnet, can be induced. This symmetry breaking can be achieved through various methods, such as Janus engineering, isovalent alloying, and external electric field. Through first-principles calculations, we start from $PT$-antiferromagnetic $\mathrm{Cr_2C_2S_6}$ and induce altermagnetic  $\mathrm{Cr_2C_2S_3Se_3}$  and fully-compensated ferrimagnetic $\mathrm{CrMoC_2S_6}$/$\mathrm{CrMoC_2S_3Se_3}$ via Janus engineering and isovalent alloying,  and the four monolayers exhibit good stabilities.
This research provides valuable insights into  transitions among net-zero-magnetization magnets, and gives an intuitive understanding of the underlying mechanisms that govern their magnetic electronic structures.

\begin{acknowledgments}
This work is supported by Natural Science Basis Research Plan in Shaanxi Province of China  (2021JM-456). We are grateful to Shanxi Supercomputing Center of China, and the calculations were performed on TianHe-2.
\end{acknowledgments}

\end{document}